\documentclass[letterpaper,conference]{IEEEtran}
\usepackage{amssymb}
\usepackage{verbatim}
\usepackage{rotating, graphicx}
\usepackage{amsfonts}
\usepackage{subfigure}
\usepackage[normalem]{ulem}
\usepackage{amsmath}
\usepackage{amsthm}
\usepackage[table]{xcolor}
\usepackage{epstopdf}
\usepackage{rotating}

\begin{document}

\title{Cooperative Coded Data Dissemination for Wireless Sensor Networks}

\author{Daniyal Ahmed and Jalaluddin Qureshi\\
Department of Electrical Engineering,
Namal College, Mianwali, Pakistan}

\maketitle
\maketitle
\begin{abstract} 
In this poster paper we present a data dissemination transmission abstraction for over the air programming (OAP) protocol which is fundamentally different from the previous hop-by-hop transmission protocols. Instead of imposing the greedy requirement that at least one node in the $i^{th}$ hop receives all packets before transmitting packets to the next hop and its neighbours, we take advantage of the spatial diversity and broadcast nature of wireless transmission to adopt a cooperative approach in which node broadcast whatever packets it has received with the expectation that it will recover the lost packets with high probability by overhearing the broadcast transmissions of its neighbours. The use of coded transmissions ensures that this does not lead to the broadcast storm problem. We validate the improved performance our of proposed transmission scheme with respect to the previous state of the art OAP protocols on a proof-of-concept two-hops TelosB wireless sensor network testbed.
\end{abstract}


\section{Introduction}
Large scale wireless sensor networks (WSN) are deployed for long period of time in isolated and harsh environment to gather information. Over such long period of time it may be necessary to update tasks which need to be performed by the motes, fix software bugs or patch security holes. Manually updating each mote for such large scale network is a challenging task. It has been reported that such reprogramming needs to be performed regularly~\cite{Dong14}. Over the air programming (OAP) protocols are protocol which enables the distribution of such code update over the network to all nodes of WSNs.

Previously proposed OAP protocols~\cite{Hui04, Dong14, Wang06, Hagedorn08} do not fully exploit spatial diversity to design a cooperative communication based transmission scheme. In this work-in-progress paper we propose a new transmission abstraction for OAP protocol which does not impose the greedy requirement that at least one node in the $i^{th}$ hop receive all the packets of a page before transmission to the next hop begins.

Due to the independent nature of successful packet reception on a wireless channel we take advantage of the observation that while a node in any $i^{th}$ hop may not necessarily receive all the $k$ packets, it is possible that the concatenation of codewords successfully received by the neighbours of a node may however result in $k$ linearly independent codewords with high probability. By adopting a suitable cooperation scheme, these nodes can cooperate amongst themselves to recover the lost packets instead of requesting the node in the previous hop to retransmit the packets.

An efficient OAP protocol should be reliable so that all the nodes of the network receive all the packets without any loss. The protocol should be scalable for network of size in order of 1,000s of nodes. The transmission protocol should take into consideration the limited hardware capacity of the motes. It should be energy efficient, as these devices are battery powered~\cite{Wang06}.

The classical approach of flooding leads to the ``broadcast storm problem''~\cite{Wang06}. In flooding every node rebroadcast the packets it has received. This scenario leads to high redundant transmissions, as a node may receive packets it has already overheard from other neighbours, and increase packet collision probability because of higher number of nodes contending to transmit.

Deluge is considered one of the seminal work on OAP protocol, and remains a \textit{de facto} OAP protocol~\cite{Hui04}. Deluge is an epidemic OAP protocol, in which each node advertises it local file object using the ADV messages. When neighbouring nodes learn that a node has more available pages it will request those pages and prepare to receive them. If the receiver loses some packets then it will request its neighbour to retransmit the lost packets using NACK-based protocol. Deluge adopts spatial multiplexing by letting nodes transmit in parallel when spaced three-hops apart to avoid interference, and enforces strict ordering of the packet requests.

However Deluge performance does not scale in lossy networks due to the large overhead of retransmissions, and ``NACK implosion problem''. In a lossy WSN, the completion time of Deluge on a 100 nodes grid topology can take upto an hour~\cite{Wang06}. To address this problem, Rateless-Deluge proposed the use of random linear codes (RLC) over field size $GF(2^8)$ to improve transmission reliability~\cite{Hagedorn08}. In RLC the source packets are linearly added to generate a codeword which is then transmitted. Upon receiving $k$ linearly independent codewords the receiver can run the Gaussian elimination decoding algorithm to recover the lost packets.

However a disadvantage of the use of RLC scheme is that the Gaussian elimination algorithm is computationally expensive. To decode $k$ codewords, the complexity is given as $\mathcal{O}(k^3+k^2L)$, where $L$ is the length of the packet, assuming multiplication tables are used to perform multiplication which has a space complexity of $q^2$, where $q$ is the field size. For decoding, the term $k^3$ is the computational complexity of performing matrix inversion using Gaussian elimination, and the term $k^2L$ is the computational complexity of multiplying the inverted matrix with the codewords. In their work it was reported that it requires 6.96 seconds on average to decode a 48 packet page on Tmote Sky mote~\cite{Hagedorn08}.

To address the high encoding and decoding computation cost of coding over $GF(2^8)$, SYNAPSE++ proposed the use of coding scheme over $GF(2)$ with sparse degree distribution \textit{partially} resembling the degree distribution of an optimized LT code~\cite{Rossi10}. The use of $GF(2)$ coding eliminates the use of multiplication table lookup, and the use of sparse degree distribution of an optimized LT code reduces the number of row addition steps needed for Gaussian elimination.

With the use of an improved coding scheme, SYNAPSE++ reported at least 6.5 improvement in decoding time compared to Rateless-Deluge based coding scheme over $GF(2^8)$. Apart from the introduction of the coding scheme the transmission abstraction adopted in Rateless-Deluge and SYNAPSE++ is similar to Deluge. 

\begin{figure}
\begin{center}
\includegraphics[scale=0.75]{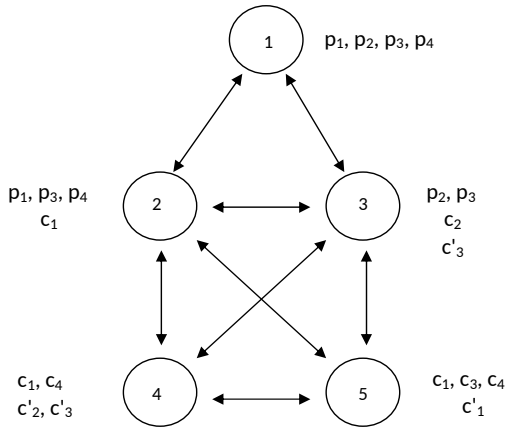}
\end{center}
\caption{Topology used for the motivating example.} \label{fig:motivating} 
\end{figure}

\section{Preliminaries}\label{sec:preliminaries}
\subsection{System Model and Notations}
Due to the limited memory size of the mote's data RAM (around 8-10KB), program code which can be in order of 10s of KB are segmented into smaller portions called pages. Each of these pages are further segmented into packets $S=[s_1, s_2, \ldots, s_k]$ of size $k$. In a random linear coding scheme, the transmitter first generates a randomly generated coding coefficient vector $G_j=[g_1, g_2, \ldots, g_k]$, $g_i\in GF(q)$, which is then multiplied with the source packets to generate the codewords as $c_j=G_j\cdot S^T$. The vector $G_j$ is included in the packet header.

Even when a relay node has not received $k$ linearly independent codewords, it can still generate new codewords $c'_i$ by recoding the received codewords~\cite{Hsu11}. Once a mote has received $k$ linearly independent codewords, decoding is performed as $H^{-1}\cdot C^T$. Where the matrix $H$, $H\in GF(q)^{k\times k}$, represents the coding coefficient matrix of the $k$ linearly independent codewords a client has received, and the matrix $C$ represents the matrix of $k$ linearly independent codewords.

\subsection{Related Work}
While the advantages of multiple-input multiple-output (MIMO) are well known towards improving the network throughput, its adaption for hardware and battery constrained devices such as smartphones and sensor motes is not practical. The use of cooperative communication where nodes ``share'' their antenna to form a virtual MIMO system, without each node having multiple antennas, has been proposed as a viable technique for such devices~\cite{Hsu11}. 

In such cooperative transmission scheme only a small subset of nodes in the networks are source(s) and receiver(s), which relaxes the transmission constraint of delivering $k$ packets to all nodes of the network as demanded in OAP protocols.

\section{Coded Cooperative Data Transmission}
\subsection{Motivating Example}
We illustrate the idea of our proposed coding scheme using a simple motivating example illustrated in Figure~\ref{fig:motivating}. We note that while the random linear coding may lead to reception of codewords which are linearly dependent, the expected number of codewords $k+\delta$ a receiver receives before decoding the $k$ codewords is given as $k+1.6$~\cite{Cruces11}. Such small redundancy of 1.6 codewords will be negligible for sufficiently large $k$, therefore without loss of generality we assume in our motivating example that it is sufficient for a receiver to receive $k$ codewords before decoding the $k$ source packets.

Node $N_1$ wishes to multicast four packets $p_1, p_2, p_3,$ and $p_4$ to node $N_2$ and $N_3$. However due to channel erasure, $p_2$ is lost at $N_2$, and $p_1$ and $p_4$ is lost at $N_3$. So none of the two nodes at second hop has all the four packets. Traditional OAP protocols would request $N_1$ to retransmit the lost packets.

However in our proposed scheme we note that, the concatenation of packet received by $N_2$ and $N_3$ results in four unique packets. So instead of adopting any error correction scheme such as requesting $N_1$ for retransmission, these two nodes can code the packets they have received and broadcast them to the nodes in next hop, node $N_4$ and $N_5$. The transmitted codewords will also be overheard by the neighbours $N_f$, $f\in\{2,3\}$. 

Assuming that $N_2$ receives $c_1$ transmitted by $N_1$ which it can use to decode $p_2$, and $N_3$ only receives $c_3$ transmitted by $N_2$. Then according to traditional approach either $N_1$ or $N_2$ should be requested to correct lost packets by $N_3$. In our case we relax this requirement. As nodes $N_4$ and $N_5$ may have received some codewords which they will broadcast to nodes in the next hop (not illustrated in this example), it is possible that codeword lost by $N_2$ can then be recovered by overhearing transmission of codewords transmitted by $N_4$ and $N_5$.

Nodes $N_4$ and $N_5$ will be generating codewords $c'_i$ by recoding the codewords $c_j$. Once $N_4$ and $N_5$ receive sufficient linearly independent codewords $c'_i$ by overhearing the transmission of one another it can then decode the $k$ source packets.

\begin{figure}
\begin{center}
\includegraphics[width=0.5\textwidth]{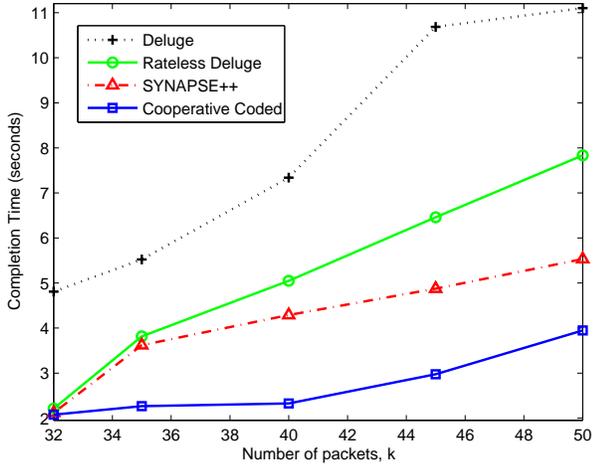}
\end{center}
\caption{Average completion time for data dissemination of $k$ packets to a two hops 5 nodes topology.} \label{fig:results} 
\end{figure}

\subsection{Implementation Details}
While we had assumed in our previous motivating example that the concatenation of codewords received by all the neighbours of a node form a full rank matrix with high probability, this may not necessarily always be the case. To address this problem we adopt the use of a request timer $\tau$. Should a node be unable to recover $k$ source packets after three transmission rounds, the first by the nodes in its previous hop, the second by the nodes in the same hop, and the third by nodes in the next hop, plus the expected time to decode those codewords, then the node request its previous hop nodes to transmit additional codewords using the NACK-protocol. The value of $\tau$ is determined based on the value of $k$.

The probability that a node does not receive $k$ linearly independent codewords after three transmission rounds would correspond to an average channel erasure probability of more than 67\% assuming codeword reception follows geometric distribution. Practical wireless channel erasure probability generally does not exceed 20-50\%.

For our OAP protocol we assume coding over $GF(2)$. Before recoding codewords, a node performs the triangularization step of Gaussian elimination to drop any linearly dependent codeword(s) it has received. By performing only the triangularization step of the Gaussian elimination instead of performing the complete decoding, we minimize the delay due to decoding before the node forwards the codewords. The use of cooperative communication provides load balancing of energy consumption to transmit codewords to next hop and can extend the lifetime of WSN.

While in this paper we allow all nodes in a given hop to participate in transmission, this may lead to collision, and redundancies at the next-hop nodes. The probability of collision with a moderate number of nodes contending to transmit on the same channel is negligible for the CC2420 radio. For upto 20 nodes contending to transmit on the same channel, the collision probability is $\leq0.01$~\cite{Dong14}. Part of our future work is to adopt a MAC transmission scheme in which only a moderate number of nodes participate in the transmission.

\subsection{Experimental Results}
We implemented our proposed transmission scheme on Advanticsys CM5000 TelosB motes running TinyOS 2.1.2. We used 5 nodes, two hops topology as show in Figure~\ref{fig:motivating} in an indoor environment. We evaluated the completion time before $k$ packets of size 20 bytes is disseminated to all nodes of the network, assuming a single page. The experiments were evaluated for Deluge, Rateless-Deluge, SYNAPSE++ and our proposed cooperative coded OAP protocol.

For each sets of parameters we repeated the experiments five times. The average completion time for each of the experiment is plotted in Figure~\ref{fig:results}. The results show that even for a simple two hop topology our proposed cooperative coded OAP protocol reduces the completion time by 80\% compared to the next best performing SYNAPSE++ OAP protocol.

\section{Conclusion} \label{sec:conclusion}
In our proposed cooperative coded transmission scheme we relaxed the requirement that it is no longer necessary for a node in the $i^{th}$ transmission to receive all packets of a page before transmitting packets to nodes in the next hop. By adopting an appropriate cooperative communication scheme based on network coding and fountain coding we proposed a cooperative coded OAP protocol. 

Testbed implementation of our proposed OAP showed significant reduction in completion time even over a simple two hops WSN. As such improvement will be additive over an increasing number of hop counts, we expect that our proposed transmission abstraction can lead to many-fold reduction in completion time for a large scale multi hop WSN, which we expect to implement as part of our future work.

\bibliographystyle{IEEEtran}
\bibliography{mainJ}

\end{document}